# Design Studies for a TPC Readout Plane Using Zigzag Patterns with Multistage GEM Detectors

B. Azmoun, P. Garg, T.K. Hemmick, M. Hohlmann, A. Kiselev, M.L. Purschke, C. Woody, A. Zhang

*Abstract*—A new Time Projection Chamber (TPC) is currently under development for the sPHENIX experiment at RHIC. The TPC will be read out using multistage GEM detectors on each end and will be divided into approximately 40 pad layers in radius. Each pad layer is required to provide a spatial resolution of ~250 microns, which must be achieved with a minimal channel count in order to minimize the overall cost of the detector. The current proposal is to make the pads into a zigzag shape in order to enhance charge sharing among neighboring pads. This will allow for the possibility to interpolate the hit position to high precision, resulting in a position resolution many times better than the 2mm pitch of the readout pads. This paper discusses various simulation studies that were carried out to optimize the size and shape of the zigzag pads for the readout board for the TPC, along with the technical challenges in fabricating it. It also describes the performance of the first prototype readout board obtained from measurements carried out in the laboratory using a highly collimated X-ray source.

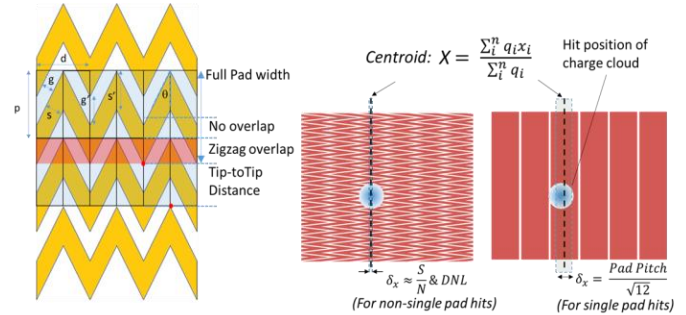

Fig. 1 The sketch on the left shows the 4 basic parameters of the zigzag pattern, including the pitch, zigzag period, gap width, and trace width, denoted by $p$, $d$, $g$, and $s$ respectively. ($\theta$, $s$', and $g$' are resultant parameters representing the characteristic angle, the trace width, and gap width at the zigzag apex.) The sketches on the right demonstrate charge sharing and centroid calculations for a zigzag and rectangular pad readout. 6 channels are shown for each pattern with a pitch of 2mm. (The drawings on the right are to scale.)

## I. INTRODUCTION

A Time Projection Chamber (TPC) is currently being built as part of the central tracking system of the sPHENIX detector at RHIC. The physics requirements demand a high performance tracking detector with a position resolution of 200-250μm per space point. However, the TPC must incorporate a relatively coarse segmentation due to cost constraints on the number of pads which can be read out. Therefore, zigzag shaped pads will be employed to enhance charge sharing among the hit pads, which serves to improve the resolution over more standard rectangular pads of equal pitch. Fig. 1 illustrates the advantage of zigzag pads in this particular application where the collected charge cloud is roughly the same size as the pitch of the readout. While the response of the zigzags is mostly governed by the signal to noise (*S/N*) ratio, similarly sized rectangular pads can be highly susceptible to single pad hits, which will severely deteriorate the overall resolution. A comparison of the performance of these two readout schemes, for a particular rectangular strip geometry and a less than optimal zigzag pattern is given in [1]. Additional studies of zigzag and chevron pad structures are given in [2], [3], [4], and [5].

## II. RESULTS

### A. Simulation

The objective of this study is to optimize the 4 parameters that define the zigzag pattern in order to optimize charge sharing among neighboring pads, while maintaining a flat response across the readout. A basic zigzag pattern was chosen since this is the simplest geometry which provides a scheme to split charge in proportion to hit position, while allowing each pad or strip to extend beyond its pitch (typically by a factor of two). Initially the pad response was studied via a rudimentary simulation, which neglects gas processes and simply allows for the uniform collection of charge onto each pad defined by the pad geometry alone. The simulated charge cloud is a 2D Gaussian distribution, whose center is scanned across the zigzag structure and at each point a charge weighted mean (or centroid) is employed to reconstruct the center of gravity. Some of the results from this simulation are shown in Fig. 2 for an ideal zigzag pattern which exhibits a linear response with minimal differential non-linearity and a spatial resolution an order magnitude better than what is implied by the physical extent of the pad itself along the position-sensitive coordinate. After studying the effects of varying the zigzag parameters, in addition to the size of the

Manuscript submitted on Jan. 5, 2018. This work was supported in part by the U.S. Department of Energy under Prime Contract No. DE-SC0012704.

B.Azmoun, A.Kiselev, M.L.Purschke, C.Woody and A.Zhang are with Brookhaven National Laboratory, Upton, NY, P.Garg and T.K.Hemmick are with Stony Brook University, Stony Brook, NY and M.Hohlmann is with Florida Institute of Technology, Melbourne, FL.

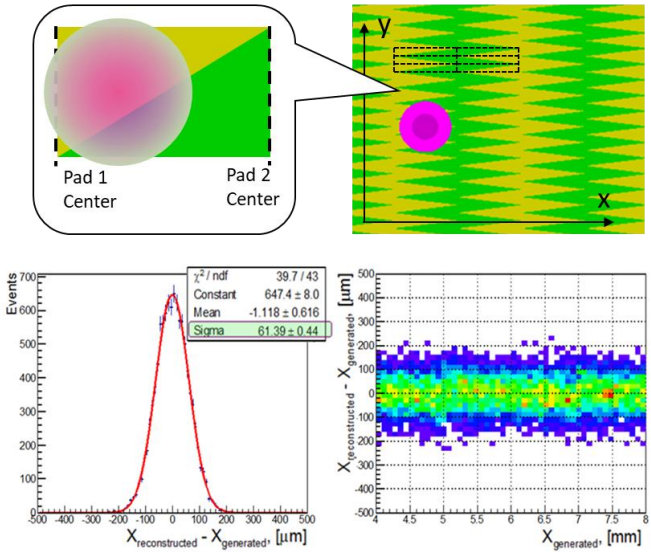

Fig. 2 Pictured on top are sketches of idealized zigzag pads with a zero space gap between them. They conceptually reveal the linear relationship between the distributed charge and the hit position. The plots on the bottom show the resultant resolution (~60μm) for a 2D Gaussian charge cloud (with $\sigma_x = \sigma_y = 400$μm) and the pad response for 2mm pitch and a 0.5mm zigzag period, which incorporates a $N/S = 2\%$.

charge cloud, several important attributes of such a linear charge sharing model were revealed, and are summarized here:

- Charge sharing is directly proportional to hit position as long as the zigzag pitch and period are chosen appropriately for the size of the charge cloud.
- The zigzag overlap should approach 100%, with minimal strip-to-strip gap width.
- Overstretching the zigzags such that the pad overlap exceeds 100% tends toward a non-linear response.
- The zigzag period should be minimized to eliminate a fluctuating response along the zigzag pattern.
- The collective response of all fired pads is independent of charge cloud size over a relatively broad range of sizes.
- The charge cloud should be mostly contained to within 2-3 pads since utilizing additional pads in the centroid calculation tends to deteriorate linearity.
- Ideally, a fired pad should not collect less charge than what is appropriate for the dynamic range and signal to noise ratio of the system.
- Ideally, the collective response of all fired pads is linearly correlated to the hit position, with a differential non-linearity (DNL) ~ 0.
- The percentage of single pad hits should be minimized to ~ 0.
- In principle there are few limiting factors for the achievable position resolution when optimized zigzag pads are employed for the readout, including practical issues like signal to noise ratio, fabrication constraints, etc.

- An optimized pad design avoids corrections for a *DNL*, which are never 100% efficient and in general depend on the charge cloud footprint size.

A separate simulation, specific to this detector application, utilizes a finite element program (ANSYS) to approximate the electric field in the induction gap of a quadruple GEM detector, responsible for carrying individual electrons of the charge cloud to the pad plane. This simulation also employs Garfield++ to take into account gas processes, including diffusion to track the trajectories of electrons onto each pad. The results in Fig. 3 illustrate the final destination of each point charge after originating from uniform discs of charge a few mm away. It is evident that despite the rather sharp points and edges of the zigzag structure, the collection of charge across the zigzag is quite uniform (at least for a 2mm gap). It should be noted that as part of the simulation, a ground plane was placed about 1/2 mm below the pad plane, sandwiching the FR4 substrate, which accounts for the fact that some charge is collected onto the gaps in between pads and lost. For this reason and in order to curb potential non-linearities, a guiding design principle is to maximize the area of the conductive layer by minimizing the gap spacing between pads.

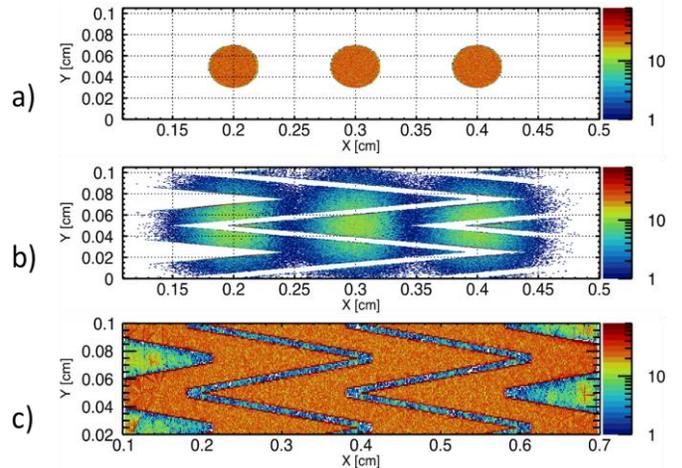

Fig. 3 a) Three discs of uniformly distributed charge arranged on a plane 2mm above the readout plane at the drift cathode. b) The density of collected electrons onto the zigzag anode pad plane after drifting across the 2mm gap and originating from the three discs. (The color scale is proportional to the charge density.) c) A plane of uniform charge that fills the acceptance is collected onto the zigzag pads for the same detector configuration. In this case the troughs of the zigzag were rounded to more accurately resemble actual manufactured electrodes.

B. Measurement

To verify the results from simulation, several readout PCB's with different zigzag patterns were studied in the lab. Each PCB made up the readout plane of a quadruple GEM detector, and Ar/$CO_2$ 70:30 was used as the working gas. The GEM detector was operated at a gain of approximately $4.5 \times 10^3$ with a drift field of 0.75 kV/cm and 3kV/cm applied to the transfer and induction gaps. The detector was studied by illuminating the acceptance with a highly collimated beam of x-rays, with a cross section of roughly 100μm by 8mm. With the x-ray source mounted to a moveable XY-stage, the collimated beam was scanned across the active area of the detector in 100μm steps, for a distance equal to several times the pad pitch along the position-sensitive coordinate. After accounting for any misalignment in the setup, the resolution for the GEM detector was calculated as the sigma of a Gaussian fit to the residual distribution. The residual was taken as the difference between the calculated centroid (using the measured charge distribution) as defined in Fig. 1, and the actual hit position, taken to be the position of the x-ray source on the XY-stage, which has an uncertainty of only a few microns. It should be noted that while the in-lab measurements of resolution may be used to attach figures of merit to each zigzag pattern, these measurements are not fully representative of the single point track resolution in the TPC, therefore the goal here is to simply maximize, to a reasonable extent this relative measure of the resolution.

Some of these results are shown below in Fig.4, 5, and 6 for a PCB recently fabricated with a zigzag pattern approaching the ideal case described above, but within the limits of what is able to be fabricated using standard chemical etching techniques. The design specifications of the zigzag pattern were for a 2mm pad pitch and ~0.5mm period, with a pad overlap of about 94%, and a conductive layer covering 67% of the active area. This level of pad overlap was chosen since there must be a comp-romise with the degree of conductor coverage for a given gap width, where the gap width was chosen as the minimal industry standard of 3mils (~75μm). While the parameters corresponding to the larger feature sizes of the design, namely the pad pitch and zigzag period were reproduced accurately, the smaller features such as the trace and gap widths were fabricated with far less accuracy. Ultimately, the generated zigzag pattern featured only 82% pad overlap and 63% conductor coverage for the actual PCB tested.

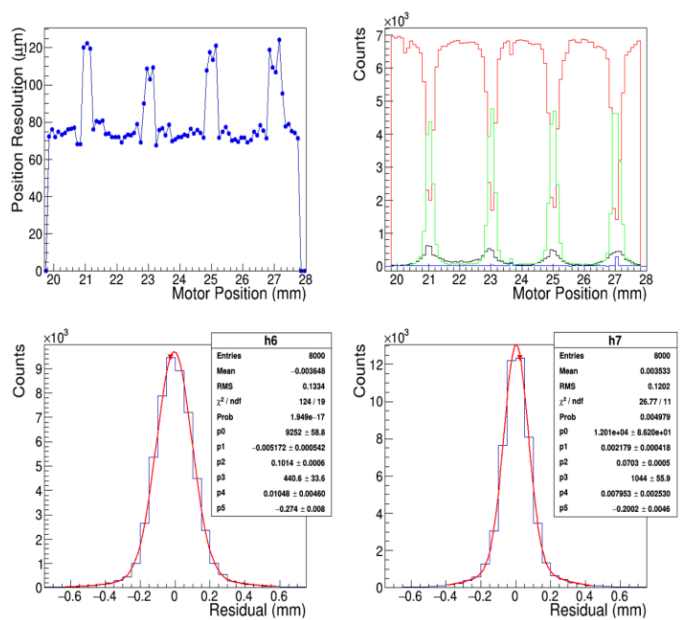

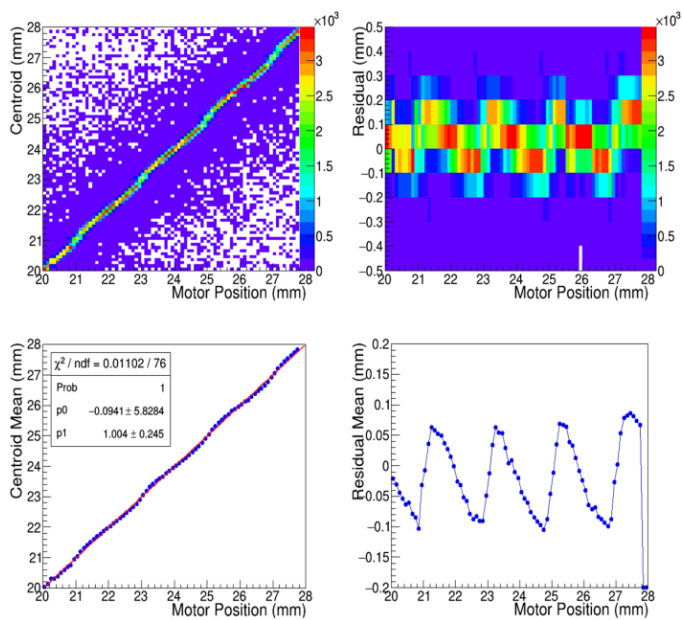

**Fig. 4** The top left scatter plot shows the reconstructed hit position vs. the motor position (i.e., the actual hit position of each x-ray modulo the width of the beam) and the right scatter plot shows the corresponding residual vs motor position. The bottom plots show the respective averages of these quantities, which more clearly depict the prevalent differential non-linearity.

**Fig. 5 Top:** position resolution and number of fired pads above threshold vs. actual hit position, respectively. For the plot on the right the black histogram corresponds to the number of single pad hits; red: 2-pad hits; green: 3-pad hits; and blue: 4-pad hits. The integrated percentage of single pad hits is only 2.6%, leaving 97.4% of the events with useful hits. **Bottom:** position residual distribution integrated over ~4 pitch cycles, without and with the DNL unfolded, respectively. The centroid is calculated using only two-pad and three-pad hits and the position resolution is taken as the sigma of the dominant Gaussian of a double Gaussian fit to each distribution.

It should be noted that while the hit position used for the centroid is only known to within the width of the x-ray beam from the collimator, no attempt was made to unfold this width from these results since the beam profile is not accurately known. However, the width is estimated to be about 40-50microns (cf. [5]), which is a significant contributor to the quoted values of resolution and should in principle be reduced by this amount subtracted in quadrature. Nevertheless, the quoted resolution from the x-ray scans are only considered relative metrics, so it is not imperative to remove a constant term that is common to all measurements.

In addition, a scan perpendicular to the sensitive coordinate (with the 8mm length of the collimator slit now parallel to the sensitive coordinate and spanning 4-5 pads) was performed to measure the full charge ratio spectrum for two pad hit events, which comprise the vast majority of all events. While the charge sharing profile is quite consistent everywhere along the zigzag period as may be expected, there is also a sharp cutoff at around 10%, implying that the minimal charge any one pad collects is most often above 10%.

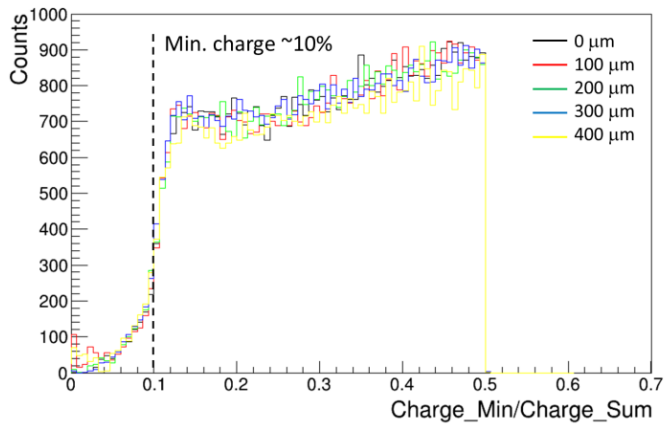

**Fig. 6 Spectra of the minimum-charge to total-charge ratio for two-pad hits for five distinct points along the zigzag period in 100μm steps. In this case, the collimator slit width spans 4-5 pads, so these results show the charge ratio integrated over several pads.**

*C. Discussion*

The PCB used for the measurements described above shows markedly improved performance compared to an earlier PCB, with the same pitch and zigzag period, but with 69% pad overlap and 66% conductor coverage. Similar x-ray scans performed with the earlier PCB, in a similar GEM setup and under similar conditions resulted in an integrated position resolution of 132μm and 98μm with and without DNL unfolding respectively, compared to 101μm and 70μm for the newer board. In addition, the number of single pad hits dropped from almost 30% to about 2.6%, as shown in Fig. 5.

The method by which the centroid was calculated used a charge threshold to identify hits to form 2, 3, and 4-pad clusters from which the centroid was derived. However, a more straight forward approach was also implemented that does not utilize a threshold cut. Rather the peak with maximum charge is identified and a contiguous three pad cluster is always formed around it by maximizing the total charge. While such a method allows for the additional influence of noise, the centroid response tends to be more flat. In this case, since the N/S was quite low, the results from the two methods were very similar, with the benefit that the "three-pad method" requires less computation. Furthermore, the three pad method also insures virtually 100% efficiency, but this may come at the cost of worse resolution in instances where charge sharing is not optimized.

The improved resolution and nearly uniform response of the newer zigzag pattern are encouraging results, which mostly validate the conclusions drawn from simulation. However, the results obtained thus far are still far from ideal, mainly due to the prevalence of a notable DNL, in addition to spikes in the position resolution and the fluctuating number of fired pads versus motor position seen in the upper plots in Fig. 5. Interestingly, the spikes in resolution coincide with sites near the center of the pads (i.e., odd numbered coordinates) where not only single pad hits are prevalent, but where charge sharing is likely weakest among two and three pad hits. In addition, the quoted resolutions are derived from double Gaussian fits (described in Fig. 5) to distributions with noticeable tails, likely from this somewhat irregular response of the readout. While these flaws in the readout performance are relatively moderate, they constitute the issues that must be contended with for the next generation zigzag based readout.

As such, the next step to realizing more ideal zigzag patterns is to overcome the distortions imposed by the fabrication process. Fig. 7 shows the two predominant distortions of the zigzag geometry due to standard chemical etching, including over-etching of the zigzag tips as well as under-etching at the troughs. These distortions put an upper limit on the pad overlap specification and contribute to further nonlinearities. Together with the limitations on the minimum gap width, the potential for implementing a linear charge sharing model in a working detector may be severely hampered by this manufacturing technique.

As a next phase of the R&D program, we are thus pursuing new fabrication processes that do not suffer from the limitations described above. In particular, the novel use of laser ablation using ultra-short and focused pulses of light to remove copper from the PCB substrate is well suited for accurately reproducing the fine detail of the zigzag structure. Such advances in the printed circuit industry will then allow one to probe regions of the zigzag geometry parameter space down to the level of 1mil (~25μm) or less.

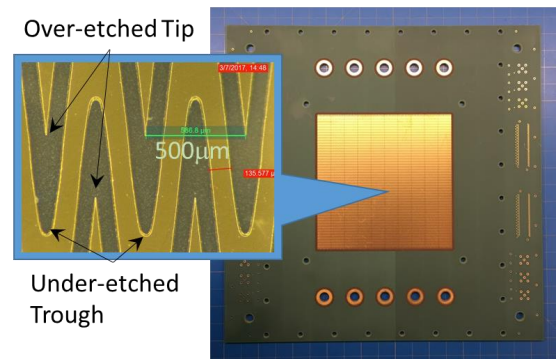

**Fig. 7 Zoomed-in features of the zigzag pattern from an actual PCB produced using standard chemical etching. The tips have been eroded by more than 125μm and the troughs under-cut by 200-300μm.**

It should be noted that there have also been preliminary attempts to compensate for the manufacturing distortions described above by implementing design features like over-stretched zigzags (cf. [5]). However, such techniques have had limited success, likely because identifying the optimal compensating design requires a dedicated effort beyond what was initially tried.

## III. Summary

With guidance from simulation we have come closer to optimizing the design of the zigzag pattern for the sense plane of a quadruple GEM readout, tailored for the sPHENIX TPC. A prototype PCB consisting of a zigzag pattern based on this design was produced and tested on the bench using collimated x-rays and showed impressive position resolution. In the process, we have also identified major limitations for the fabrication of the zigzag pads, which severely limits the potential of the ideal pad designs. However, we believe that by overcoming such limitations, the use of zigzag pads can be a broadly reaching alternative to many highly segmented, and therefore very costly readout options.